\documentclass[showpacs,aps,graphicx,twocolumn]{revtex4}

\usepackage{amsmath}
\usepackage{graphicx}
\usepackage{txfonts}

\begin{document}


\title{Physically feasible three-level transitionless quantum driving with multiple Schr\"{o}dinger dynamics\footnote{Published in Phys. Rev. A \textbf{93}, 052324 (2016)}}
\author{Xue-Ke Song, Qing Ai, Jing Qiu, and Fu-Guo Deng\footnote{Corresponding author: fgdeng@bnu.edu.cn} }
\address{Department of Physics, Applied Optics Beijing Area Major Laboratory,
Beijing Normal University, Beijing 100875, China }
\date{\today }

\begin{abstract}

Three-level quantum systems, which possess some unique
characteristics beyond two-level ones, such as electromagnetically
induced transparency, coherent trapping, and Raman scatting, play
important roles in solid-state quantum information processing. Here,
we introduce an approach to implement the physically feasible
three-level transitionless quantum driving with multiple
Schr\"{o}dinger dynamics (MSDs). It can be used to control
accurately population transfer and entanglement generation for
three-level quantum systems in a nonadiabatic way.  Moreover, we
propose an experimentally realizable hybrid architecture, based on
two nitrogen-vacancy-center ensembles coupled to a transmission line
resonator, to realize our transitionless scheme which requires fewer
physical resources and simple procedures, and it is more robust
against environmental noises and control parameter variations than
conventional adiabatic passage techniques. All these features
inspire the further application of MSDs on robust quantum
information processing in experiment.
\end{abstract}
 \maketitle


\section{Introduction}
\label{sec1}

Accurately controlling a quantum system with high fidelity is a
fundamental prerequisite in quantum information processing
\cite{QIP1}, high precision measurements \cite{precision
measurement}, and coherent control of atomic and molecular systems
\cite{molecular1}. To this end, rapid adiabatic passage \cite{rap},
which leads two-level quantum systems to evolve slowly enough along
a specific path, can produce near-perfect population transfer
between two quantum states of (artificial) atoms or molecules. The
adiabatic evolution requires long runtime, which will generate the
extra loss of coherence and spontaneous emission of quantum systems.
Shortcuts to adiabaticity are alternative fast processes to
reproduce the same physical processes in a finite shorter time,
which is only limited by the energy-time complementarity \cite{qst}.
There are two potentially equivalent shortcuts to speed up adiabatic
process in a nonadiabatic route: Lewis-Riesenfeld invariant-based
inverse engineering \cite{LR1,LR2,LR3,LR4,STAreview} and
transitionless quantum driving (TQD)
\cite{TQDA1,Rice,TQDA2,TQDA3,TQDA5,TQDA6,dr}. Interestingly, TQD has
attracted considerable attention in experiment
 \cite{TQDAexperiment,TQDANV}. In 2012, Bason \emph{et al.} \cite{TQDAexperiment}
demonstrated the quantum system following the instantaneous
adiabatic ground state nearly perfectly on Bose-Einstein condensates
in optical lattices. In 2013, Zhang \emph{et al.} \cite{TQDANV}
implemented the assisted adiabatic passages through TQD in a
two-level quantum system by controlling a single spin in an
nitrogen-vacancy center in diamond.

For three-level quantum systems, the stimulated Raman adiabatic
passage (STIRAP) technique \cite{STIRAP} uses partially overlapping
pulses (Stokes and pump pulses) to perfectly realize the population
transfer between two quantum states with the same parity, in which
single-photon transitions are forbidden by electric dipole
radiation. The STIRAP over the rapid adiabatic passage is its
robustness against substantial fluctuations of pulse parameters,
since the evolution of the quantum system is in the dark state space
and only the two quantum states are involved. This technique has
gained theoretical and experimental studies in atomic and molecular
\cite{shore} and superconducting quantum systems \cite{paraoanu}.
When TQD is applied to speed up the adiabatic operation in
three-level quantum systems, the situation becomes more complicated
\cite{TQDA3,Garaot,TQDA7,TQDA8,wei}. In 2010, Chen \emph{et al.}
\cite{TQDA3} employed the TQD to speed up adiabatic passage
techniques in three-level atoms extending to the short-time domain
their robustness with respect to parameter variations. In 2014,
Mart\'{\i}nez-Garaot \emph{et al.} \cite{Garaot} studied shortcuts
to adiabaticity in three-level systems by means of Lie transforms.
Alternatively, in 2012, Chen and Muga \cite{chen} designed the
resonant laser pulses to perform the fast population transfer in
three-level systems by invariant-based inverse engineering. In 2014,
Kiely and Ruschhaupt \cite{ruschhaupt} constructed fast and stable
control schemes for two- and three-level quantum systems.

Interestingly, multiple Schr\"{o}dinger dynamics (MSDs)
\cite{ibanez1,ibanez2} were presented to adopt iterative
interaction pictures to get physically feasible interactions or
dynamics for two-level quantum systems recently. Meanwhile, it
enables the designed interaction picture to reproduce the same final
population (or state) as those in the original Schr\"{o}dinger
picture by appropriate boundary conditions. In 2012,
Ib\'{a}\~{n}ez \emph{et al.} \cite{ibanez1} first employed several
Schr\"{o}dinger pictures and dynamics to design alternative and
feasible experimental routes for trap expansions and compressions,
and for harmonic transport. In 2013, Ib\'{a}\~{n}ez \emph{et al.}
\cite{ibanez2} also examined the limitations and capabilities of
superadiabatic iterations to produce a sequence of shortcuts to
adiabaticity by iterative interaction pictures.  This raises a
significative question: whether one can find an effective way for
three-level TQD in experimental applications. Three-level quantum
systems play important roles in  solid-state quantum information
processing as they possess some unique characteristics beyond
two-level ones, such as electromagnetically induced transparency,
coherent trapping, Raman scatting, and so on. Therefore,
manipulating such quantum systems in an accurate and robust manner
is especially important.

Inspired by the two-level TQD with MSDs \cite{ibanez1,ibanez2}, here we
employ the iteration process to obtain physically
feasible TQD in three-level quantum systems. More interestingly, we present a
physical implementation for the transitionless scheme with the
hybrid quantum system composed of nitrogen-vacancy-center ensembles (NVEs)
and the superconducting transmission line resonator (TLR). It has some advantages.
First, it can accurately control quantum systems in a shorter time, as adiabatic
quantum evolution can be efficiently accelerated by TQD. Second, the MSDs-based Hamiltonian
required for three-level TQD is physically feasible, which can be used to
implement accurate and robust population transfer and entanglement
generation with high fidelity in a single-shot operation. Third, it
is more robust against control parameter fluctuations and
dissipations than conventional adiabatic passage technique. Fourth,
the transitionless scheme presented here is quite universal, and
it is broadly applicable in other quantum systems, such as atom cavity,
superconducting-qubit TLR, and so on. All
these advantages provide the good applications of MSDs on robust
quantum information processing in experiment in the future.

This paper is organized as follows: In Sec. \ref{sec2}, we show the
basic principle of our scheme for obtaining the physically feasible
TQD in three-level quantum systems by using MSDs. In Sec.\ref{sec3},
we give specific comparisons of population transfer and
superposition state generation based on the conventional STIRAP and
MSDs, respectively. In Sec. \ref{sec4}, we present a physical
implementation of the transitionless scheme on an NVEs-TLR system,
and analyze the fidelity in the presence of decoherence.  A
discussion and a summary are enclosed in Sec. \ref{sec5}.

\section{Physically feasible Hamiltonian with MSDs on three-level quantum systems}
\label{sec2}

\subsection{The transitionless Hamiltonian with multiple Schr\"{o}dinger dynamics}
\label{sec21}

First, we give a brief review of TQD. Considering an arbitrary
time-dependent Hamiltonian $H_{0}(t)$ of a quantum system, which has
the nondegenerate instantaneous eigenstates $|n_{0}(t)\rangle$ with
corresponding eigenvalues $E_{n}(t)$, we get
\begin{eqnarray}    
H_{0}(t)|n_{0}(t)\rangle=E_{n}(t)|n_{0}(t)\rangle.
\end{eqnarray}
In the adiabatic approximation, the state evolution of the system
driven by $H_{0}(t)$ can be written as ($\hbar =1$)
\begin{eqnarray}    
|\Psi_{n}(t)\rangle=exp\left\{-i\!\!\int_{0}^{t}\!\!dt^{'}\!E_{n}(t^{'})
-\!\!\int_{0}^{t}\!\!dt^{'}\langle n_{0}(t^{'})|\dot{n_{0}}(t^{'})\rangle\right\}|n_{0}(t)\rangle.
\end{eqnarray}
As a consequence, the evolution operator for this given quantum
system is specified. Alternatively, one can seek a transitionless
Hamiltonian $H(t)$ that can accurately drive
evolving state $|\psi_{n}(t)\rangle$ in a shortest possible
time, which guarantees that there are no transitions between
the eigenstates of $H_{0}(t)$. That is, it should satisfy
\begin{eqnarray}    
H(t)|\psi_{n}(t)\rangle=i|\dot{\psi_{n}}(t)\rangle.
\end{eqnarray}
Defining a time-dependent unitary operator
\begin{eqnarray}    
\!\!U(t)\!=\!\!\!\sum_{n}\!exp\!\left\{\!-i\!\!\int_{0}^{t}\!\!dt^{'}\!E_{n}(t^{'})
-\!\!\int_{0}^{t}\!\!dt^{'}\langle
n_{0}(t^{'})|\dot{n_{0}}(t^{'})\rangle\!\right\}\!|n_{0}(t)\rangle\langle
n_{0}(0)|,\nonumber\\
\!\!\!\!\!\!
\end{eqnarray}
which obeys $H(t)U(t)=i\dot{U}(t)$. By analytically solving
the equation $H(t)=i\dot{U}(t)U^{\dag}(t)$, we have
\begin{eqnarray}    
H(t)&=&\sum_{n}E_{n}|n_{0}\rangle\langle n_{0}|+i\sum_{n}(|\dot{n_{0}}\rangle \langle n_{0}|
-\langle n_{0}|\dot{n_{0}}\rangle|n_{0}\rangle\langle n_{0}|)  \nonumber\\
&\equiv&H_{0}(t)+H^{cd}_{0}(t), \label{shortcuts}
\end{eqnarray}
where all kets are time-dependent. From Eq.~(\ref{shortcuts}), one
can see that the transitionless Hamiltonian $H(t)$ consists of the
original Hamiltonian $H_{0}(t)$ for adiabatic evolution and a
counterdiabatic driving Hamiltonian $H^{cd}_{0}(t)$ \cite{TQDA1,
Rice,TQDA2,TQDA3}. TQD offers an effective accurate route for the
controlled system following perfectly the instantaneous ground state
of a given Hamiltonian in theory and experiment. Nevertheless, it is
found that the transitionless Hamiltonian is difficult to
implement for example in three-level quantum systems \cite{TQDA3} since the
counterdiabatic driving Hamiltonian has to break down the energy structure of the
original Hamiltonian or bring extra detunings.

Superadiabatic iterations as an extension of the usual adiabatic
approximation have been introduced in Ref.~\cite{berry}.
The process of superadiabatic iteration
can be summarized in Table \ref{table1}, where $H_{j}(t)$ donates
the $j-th$ Hamiltonian by a unitary transformation $A_{j-1}(t)$ on
the $(j-1)-th$ Hamiltonian $H_{j-1}(t)$,
$A_{j}(t)=\sum\limits_{n}|n_{j}(t)\rangle\langle n_{j}(0)|$
($j=1,2,\cdot\cdot\cdot$), $|n_{j}(t)\rangle$ are the eigenstates of
the Hamiltonian $H_{j}(t)$, and $j$ is the number of superadiabatic
iteration.

\begin{table}[htb]
\centering \caption{Scheme for superadiabatic iteration to realize
MSDs as a result of  Hamiltonian $H_{0}(t)$.}
\begin{tabular}{cccc}
\hline\hline
Iteration&            Hamiltonian                  & Eigenstates                  &  Unitary operator     \\
   \hline
$0-th$   &         $H_{0}(t)$         &  $|n_{0}(t)\rangle$           & $A_{0}(t)\!=\!\sum\limits_{n}|n_{0}(t)\rangle\langle n_{0}(0)|$ \\
$1-st$   &        $H_{1}(t)$         & $|n_{1}(t)\rangle$           & $A_{1}(t)\!=\!\sum\limits_{n}|n_{1}(t)\rangle\langle n_{1}(0)|$ \\
$\cdot\cdot\cdot\cdot\cdot\cdot$&&&\\
$j-th$   &         $H_{j}(t)$         &  $|n_{j}(t)\rangle$           & $A_{j}(t)\!=\!\sum\limits_{n}|n_{j}(t)\rangle\langle n_{j}(0)|$ \\
$\cdot\cdot\cdot\cdot\cdot\cdot$&&&\\
\hline\hline
\end{tabular}\label{table1}
\end{table}

Here, our goal is to use MSDs to obtain physically
feasible transitionless Hamiltonian for three-level quantum systems
in TQD. In what follows we will present an explicit explanation
about it, reviewing the ideas from Refs.~\cite{ibanez1,ibanez2}.
For the initial Hamiltonian $H_{0}(t)$ with eigenstates $|n_{0}(t)\rangle$,
the corresponding
transitionless Hamiltonian $H^{T}_{0}(t)$ for $0\!-\!th$ iteration
reads
\begin{eqnarray}    
H^{T}_{0}(t)=H_{0}(t)+H^{cd}_{0}(t)=\sum_{n}E_{n}|n_{0}\rangle\langle n_{0}|+i\dot{A}_{0}(t)A^{\dag}_{0}(t),
\end{eqnarray}
where $A_{0}(t)=\sum\limits_{n}|n_{0}(t)\rangle\langle n_{0}(0)|$ is
defined as the unitary operator based on the eigenstates
$|n_{0}(t)\rangle$ of the initial Hamiltonian $H_{0}(t)$, and
$|n_{0}(0)\rangle$ is the bare adiabatic basis. Here, the
eigenstates are chosen to fulfill the parallel transport condition,
i.e., $\langle n_{0}(t)|\dot{n}_{0}(t)\rangle=0$.

In the first interaction picture (the $1\!-\!th$ iteration), by
a unitary transformation $A_{0}(t)$, the interaction picture Hamiltonian $H_{1}(t)$
becomes
\begin{eqnarray}    
H_{1}(t)=A^{\dag}_{0}(t)[H_{0}(t)-K_{0}(t)]A_{0}(t).\;\;\;\;\;\;
\label{IP}
\end{eqnarray}
where $K_{0}(t)=i\dot{A}_{0}(t)A^{\dag}_{0}(t)$. In this case, the
transitionless Hamiltonian is described by
\begin{eqnarray}    
H^{T}_{1}(t)&=&A^{\dag}_{0}(t)[H_{0}(t)\!-\!K_{0}(t)\!+\!H^{cd}_{0}(t)]A_0(t)
\nonumber\\ &=&A^{\dag}_{0}(t)H_{0}(t)A_{0}(t).\;\;
\end{eqnarray}
Here, we employ the relation $K_{0}(t)\!=\!H^{cd}_{0}(t)$. In the
Schr\"{o}dinger picture, the Hamiltonian for TQD
is $H_{0}(t)+H^{cd}_{0}(t)$. It is worth noticing that
$H^{T}_{0}(t)$ and $H^{T}_{1}(t)$ are related by a unitary transform
$A_{0}(t)$, and they represent the same common underlying physics.

In the second interaction picture (the $2\!-\!th$ iteration), for the Hamiltonian
$H_{1}(t)$ with eigenstates $|n_{1}(t)\rangle$, the interaction picture Hamiltonian
$H_{2}(t)$ can be expressed as
\begin{eqnarray}    
H_{2}(t)=A^{\dag}_{1}(t)[H_{1}(t)-K_{1}(t)]A_{1}(t),
\end{eqnarray}
where $A_{1}(t)=\sum\limits_{n}|n_{1}(t)\rangle\langle n_{1}(0)|$
and $K_{1}(t)=i\dot{A}_{1}(t)A^{\dag}_{1}(t)$. In the same way, by
adding a counterdiabatic driving term, one can obtain another
transitionless Hamiltonian $H^{T}_{2}(t)$. Then the Hamiltonian in TQD
is $H_{0}(t)+H^{cd}_{1}(t)$, where
$H^{cd}_{1}(t)=A_{0}(t)K_{1}(t)A^{\dag}_{0}(t)$.

Similarly, in the high-order interaction picture [the $(j+1)-th$
iteration], one can also get the corresponding Hamiltonian to
realize TQD in the Schr\"{o}dinger picture as
\begin{eqnarray}    
H_{0}(t)+H^{cd}_{j}(t)=H_{0}(t)+iB_{j}(t)\dot{A}_{j}(t)A^{\dag}_{j}(t)B^{\dag}_{j}(t),
\;\;\;\;\;\label{CD}
\end{eqnarray}
where $B_{j}(t)=A_{0}(t)A_{1}(t)\cdot\cdot\cdot A_{j-1}(t)$  and
$A_{j}(t)=\sum\limits_{n}|n_{j}(t)\rangle\langle n_{j}(0)|$
($j=1,2,\cdot\cdot\cdot$) with $|n_{j}(t)\rangle$ being the
eigenstates of the Hamiltonian $H_{j}(t)$ for the $j-th$
iteration. Note that a physically feasible Hamiltonian is hard to
obtain due to the unpredictable number of superadiabatic
iterations needed for execution in the specific quantum systems.

\subsection{Physically feasible three-level transitionless quantum driving}
\label{sec22}

In three-level quantum systems, the effective Hamiltonian for
achieving adiabatic population transfer in the orthogonal basis of
$\{|\phi_{1}\rangle, |\phi_{2}\rangle, |\phi_{3}\rangle\}$ takes the form of
\begin{eqnarray} 
H_{0}(t) & = \eta\left(
              \begin{array}{ccc}
                0           & 0           & \cos\theta \\
                0           & 0           & \sin\theta \\
                \cos\theta  & \sin\theta  & 0 \\
              \end{array}
            \right),\label{effective}
\end{eqnarray}
where $\eta=\sqrt{\eta_{1}^2+\eta_{2}^2}$,
$\theta=\arctan(\eta_{1}/\eta_{2})$, and $\eta$, $\eta_{1}$, and
$\eta_{2}$ are time-dependent effective coupling strengths. The
instantaneous eigenvalues and the corresponding normalized
eigenstates are
\begin{align} 
\begin{split}
E_{\mp}&=\mp\eta, \;\;\;\;\;E_{0}=0,\\
|E_{-}\rangle&=\frac{1}{\sqrt{2}}(\cos\theta|\phi_{1}\rangle+\sin\theta|\phi_{2}\rangle-|\phi_{3}\rangle),\\
|E_{+}\rangle&=\frac{1}{\sqrt{2}}(\cos\theta|\phi_{1}\rangle+\sin\theta|\phi_{2}\rangle+|\phi_{3}\rangle),\\
|E_{0}\rangle&=\sin\theta|\phi_{1}\rangle-\cos\theta|\phi_{2}\rangle.
\label{dark}
\end{split}
\end{align}
It is easy to see that $\langle E_{m}|\dot{E}_{m}\rangle=0$
$({m}={+},{-},{0})$. From Eq.~(\ref{IP}), one can obtain the interaction picture
Hamiltonian in the $1\!-\!th$ iteration for the effective
Hamiltonian $H_{0}(t)$ in the basis $\{|E_{-}\rangle, |E_{+}\rangle,
|E_{0}\rangle\}$ as follows:
\begin{eqnarray} 
H_{1}(t) & = \left(
              \begin{array}{ccc}
                -\eta                           & 0                               & -\frac{i\dot{\theta}}{\sqrt{2}} \\
                0                               &\eta                             & -\frac{i\dot{\theta}}{\sqrt{2}} \\
                \frac{i\dot{\theta}}{\sqrt{2}}  & \frac{i\dot{\theta}}{\sqrt{2}}  & 0 \\
              \end{array}
            \right),
\end{eqnarray}
where
$\dot{\theta}=(\dot{\eta}_{1}\eta_{2}-\dot{\eta}_{2}\eta_{1})/\eta^2$.
The unitary transform matrix related to $H_{0}(t)$ and $H_{1}(t)$ is
\begin{eqnarray} 
A_{0} & = \left(
              \begin{array}{ccc}
                \frac{1}{\sqrt{2}}\cos\theta    & \frac{1}{\sqrt{2}}\cos\theta               & \sin\theta  \\
                \frac{1}{\sqrt{2}}\sin\theta    &\frac{1}{\sqrt{2}}\sin\theta                & -\cos\theta  \\
                -\frac{1}{\sqrt{2}}             & \frac{1}{\sqrt{2}}                         & 0 \\
              \end{array}
            \right). \label{A0}
\end{eqnarray}
The normalized eigenvectors of the Hamiltonian $H_{1}(t)$ are
\begin{align} 
\begin{split}
\lambda_{\mp}&=\mp\sqrt{\eta^2+\dot{\theta}^2}, \;\;\;\;\;\lambda_{0}=0,\\
|\lambda_{-}\rangle&=\frac{iW}{R}|E_{-}\rangle+\frac{iQ}{R}|E_{+}\rangle+\frac{\sqrt{2}\dot{\theta}}{R}|E_{0}\rangle,\\
|\lambda_{+}\rangle&=\frac{iQ}{R}|E_{-}\rangle+\frac{iW}{R}|E_{+}\rangle-\frac{\sqrt{2}\dot{\theta}}{R}|E_{0}\rangle,\\
|\lambda_{0}\rangle&=\frac{-i\sqrt{2}\dot{\theta}}{R}|E_{-}\rangle+\frac{i\sqrt{2}\dot{\theta}}{R}|E_{+}\rangle+\frac{2\eta}{R}|E_{0}\rangle,
\end{split}
\end{align}
where $W=\eta+\sqrt{\eta^2+\dot{\theta}^2}$,
$Q=-\eta+\sqrt{\eta^2+\dot{\theta}^2}$, and
$R=2\sqrt{\eta^2+\dot{\theta}^2}$. It generates the unitary operator
\begin{eqnarray} 
A_{1} & = \left(
              \begin{array}{ccc}
                \frac{iW}{R}                         & \frac{iQ}{R}                        & \frac{-i\sqrt{2}\dot{\theta}}{R}  \\
                \frac{iQ}{R}                         &\frac{iW}{R}                         & \frac{i\sqrt{2}\dot{\theta}}{R} \\
                \frac{\sqrt{2}\dot{\theta}}{R}       & -\frac{\sqrt{2}\dot{\theta}}{R}     & \frac{2\eta}{R} \\
              \end{array}
            \right), \label{A1}
\end{eqnarray}
with which one can get the interaction picture Hamiltonian in the $2\!-\!th$
iteration. Substituting Eq.~(\ref{A0}) and Eq.~(\ref{A1}) into
Eq.~(\ref{CD}) when $j=1$, one can obtain the Hamiltonian $H_{M}(t)$
in MSDs for realizing shortcuts to adiabaticity as
\begin{eqnarray} 
H_{M}(t) & = \left(
              \begin{array}{ccc}
                0                                         & 0                                       &\!\! \eta\!\cos\!\theta+V_{1} \\
                0                                         & 0                                         &\!\! \eta\sin\!\theta-V_{2} \\
                \eta\!\cos\!\theta+V_{1}  & \eta\!\sin\theta-V_{2}  & 0 \\
              \end{array} \label{Hamiltonian}
            \right),
\end{eqnarray}
where $V_{1}=4\sin\!\theta(\dot{\eta}\dot{\theta}-\eta\ddot{\theta})/R^2$,
$V_{2}=4\cos\!\theta(\dot{\eta}\dot{\theta}-\eta\ddot{\theta})/R^2$,
$\dot{\eta}=(\eta_{1}\dot{\eta_{1}}+\eta_{2}\dot{\eta_{2}})/\eta$,
and
$\ddot{\theta}=[(\ddot{\eta_{1}}\eta_{2}-\eta_{1}\ddot{\eta_{2}})\eta
-2\dot{\eta}(\dot{\eta_{1}}\eta_{2}-\eta_{1}\dot{\eta_{2}})]/\eta^3$.
It is not difficult to find
that the Hamiltonian $H_{M}(t)$ in MSDs has the same form
as the Hamiltonian $H_{0}(t)$, without additional couplings and
detunings. Thus, a simple and feasible control of TQD for
three-level systems is physically implemented with MSDs by flexibly
tuning the effective coupling strengths.

\begin{figure}[!h]
\begin{center}
\includegraphics[width=8.0 cm,angle=0]{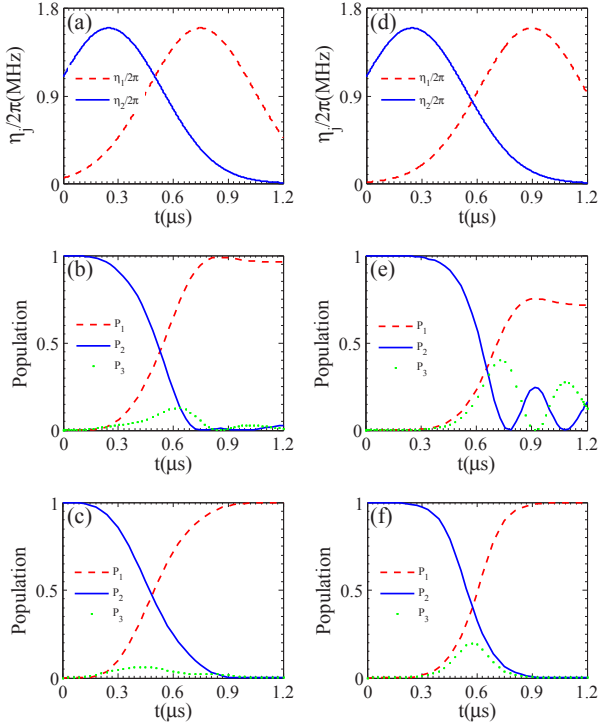}
\caption{ Comparisons of robustness of the population
transfer based on the STIRAP and  MSDs. (a) The time-dependent
effective coupling strengths $\eta_{1}$ and $\eta_{2}$ are in the
Gaussian shapes as $\eta_{j}=\eta_{0}e^{-[(t-t_{j})/T]^2}$ ($j=1,2$)
with $\eta_{0}/{2\pi}=1.6~\rm MHz$, $t_{1}=0.75~\rm\mu s$,
$t_{2}=0.25~\rm\mu s$, and $T=0.408~\rm\mu s$. Time evolution of the
population $P_{k}(t)$ during the population transfer from
$|\phi_{2}\rangle$ to $|\phi_{1}\rangle$ based on: (b) STIRAP, and
(c) MSDs, respectively. When the time delay of $\eta_{1}$ is changed
to be $t_{1}=0.9~\rm\mu s$ and other parameters remain invariant,
time evolution of the effective coupling strengths and the
populations based on STIRAP and  MSDs
 are shown in (d), (e), (f), respectively.}\label{fig1}
\end{center}
\end{figure}

\section{High-fidelity population transfer and superposition state generation} \label{sec3}

\subsection{Population transfer} \label{sec31}

From Eq.~(\ref{dark}), one can see that when $\theta=0$ at time
$t_{i}=0$, the dark state $|E_{0}(t)\rangle$ becomes
$|\phi_{2}\rangle$ with a global phase factor $\pi$. If the system
evolves adiabatically along the state $|E_{0}(t)\rangle$, the final
state is $|\phi_{1}\rangle$ when $\theta=\frac{\pi}{2}$ at later
time $t_{f}$. As a result, a simple population transfer is
completely realized by STIRAP \cite{STIRAP}. For this purpose, the
time-dependent effective coupling strengths are in the Gaussian
shapes as
\begin{eqnarray}    
\eta_{j}=\eta_{0}e^{-[(t-t_{j})/T]^2}, \label{driving}
\end{eqnarray}
where $\eta_{0}$, $t_{j}$, and $T$ are the amplitude, time delay,
and width of the coupling strength, respectively. In Fig.
\ref{fig1}(a), we display variations of the two optimal effective
coupling strengths with time $t$ for achieving population transfer,
where $\eta_{0}/{2\pi}=1.6~\rm MHz$, $t_{1}=0.75~\rm\mu s$,
$t_{2}=0.25~\rm\mu s$, and $T=0.408~\rm\mu s$. Figures. \ref{fig1}(b)
and \ref{fig1}(c) present time evolution of the populations
during the transfer process from $|\phi_{2}\rangle$ to
$|\phi_{1}\rangle$ based on STIRAP and  MSDs, respectively. The
population is defined as
$P_{k}(t)=\langle\phi_{k}|\rho(t)|\phi_{k}\rangle$ ($k=1,2,3$) with
$\rho(t)$ being the time evolution of density matrix after the
population transfer operation on the initial state
$|\phi_{2}\rangle$. In this case, both the time evolutions governed
by the Hamiltonians $H_{0}(t)$ and $H_{M}(t)$ can achieve near-perfect
population transfer from $|\phi_{2}\rangle$ to $|\phi_{1}\rangle$,
while the population $P_{3}$ of intermediate state
$|\phi_{3}\rangle$ shows a slightly different behavior. When the
time delay of $\eta_{1}$ is changed to be $t_{1}=0.9~\rm\mu s$,
which reduces overlap of the two effective coupling strengths, we
plot variations of the effective coupling strengths, time evolution
of the populations based on STIRAP and MSDs in Figs. \ref{fig1}(d),
\ref{fig1}(e), and \ref{fig1}(f), respectively. One can see that the population transfer by
the Hamiltonian $H_{M}(t)$ with MSDs is perfectly realized in a short
evolution time, and the final population of the target state
$|\phi_{1}\rangle$ can reach $100\%$, while the Hamiltonian
$H_{0}(t)$ with STIRAP cannot. Moreover, numerical calculations
reveal that the Hamiltonian $H_{M}(t)$ is also valid for high-fidelity
population transfer even the time delay of $\eta_{1}$ becomes much
bigger than $0.9~\rm\mu s$, suggesting that our transitionless
scheme with MSDs is very robust and can efficiently realize perfect
population transfer.

\begin{figure}[tpb]
\begin{center}
\includegraphics[width=8 cm,angle=0]{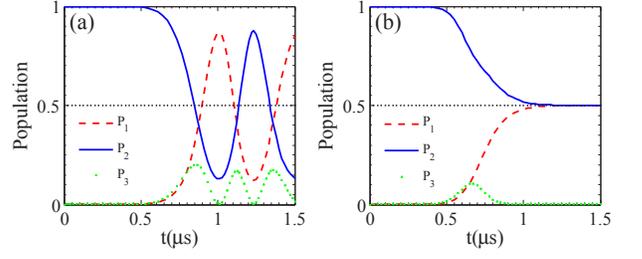}
\caption{ Time evolutions of the populations in the
superposition state generation scheme for
$|\phi_{1}\rangle$, $|\phi_{2}\rangle$, and $|\phi_{3}\rangle$ based
on (a) the STIRAP and (b) MSDs, respectively, with
$\eta_{0}/{2\pi}=1.6~\rm MHz$, $t_{3}=1.15~\rm\mu s$,
$t_{4}=0.25~\rm\mu s$, and $T=0.408~\rm\mu s$.}\label{fig2}
\end{center}
\end{figure}

\subsection{Superposition state generation}
\label{sec32}

Assuming the initial state of the system is $|\phi_{2}\rangle$, one
can easily get the superposition state
$|\psi\rangle=\frac{1}{\sqrt{2}}(|\phi_{1}\rangle-|\phi_{2}\rangle)$
by STIRAP and MSDs. For this purpose, the two time-dependent
effective coupling strengths are designed as
\begin{align}
\begin{split}   
\eta_{1}&=\eta_{0}e^{-[(t-t_{3})/T]^2},\\
\eta_{2}&=\eta_{0}e^{-[(t-t_{4})/T]^2}+\eta_{0}e^{-[(t-t_{3})/T]^2},
\end{split}
\end{align}
which should satisfy the boundary conditions of the STIRAP that at
the beginning of the operation $\eta_{1}/\eta_{2}=0$ and at the end
$\eta_{1}/\eta_{2}=1$. Given the parameters $\eta_{0}/{2\pi}=1.6~\rm
MHz$, $t_{4}=0.25~\rm\mu s$, and $T=0.408~\rm\mu s$, the
performances of the populations for $|\phi_{1}\rangle$ and
$|\phi_{2}\rangle$ with variation of $t_{3}$ have two conditions as
follows: $(\rm\expandafter{\romannumeral1})$ when the parameter
$t_{3}$ gets an optimal time $0.75~\rm\mu s$, time evolutions of the
populations $P_{1}$ and $P_{2}$ with STIRAP and MSDs reach an
approximate value $\frac{1}{2}$, that is, the two approaches
effectively generate the superposition state $|\psi\rangle$;
$(\rm\expandafter{\romannumeral2})$ when $t_{3}$ increases, the
population dynamics with STIRAP and MSDs exhibit significantly
different behaviors. The equivalent populations with
$P_{1}=P_{2}=\frac{1}{2}$ can be implemented with MSDs, implying
that time evolution of the quantum state governed by  $H_{M}(t)$ is in
the superposition state $|\psi\rangle$, while the Hamiltonian
$H_{0}(t)$ in STIRAP leads to oscillatory behaviors for $P_{1}$ and
$P_{2}$, as shown in Figs. \ref{fig2}(a) and \ref{fig2}(b), respectively. These results
convince us that MSDs could pave an efficient
way to achieve accurate and robust quantum information processing.

\begin{figure}[tpb]
\begin{center}
\includegraphics[width=8 cm,angle=0]{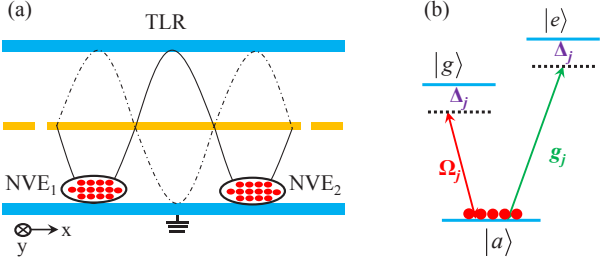}
\caption{(a) Schematic diagram of the hybrid quantum
system, which consists of two NVEs coupled to a high-Q TLR. (b) The
$V$-type energy-level configuration for the ground state of NVE
driven by the resonator and appropriate external magnetic fields.
}\label{fig3}
\end{center}
\end{figure}

\section{physical implementation of the transitionless
scheme on an NVEs-TLR system} \label{sec4}

To experimentally realize the population transfer and entanglement
generation, we consider the hybrid quantum system, in which two NVEs
are coupled to a high-Q TLR, as shown in Fig. \ref{fig3}(a). The NVE
can be modeled as a $V$-style three-level qubit with $|g\rangle$ and
$|e\rangle$ being two upper-levels, and $|a\rangle$ serving as the
lower-level. As illustrated in Fig. \ref{fig3}(b), the transition
$|a\rangle\leftrightarrow|e\rangle$ is largely detuned to the
resonator frequency with coupling strength $g_{j}$ and detuning
$\Delta_{j}$, and the transition $|a\rangle\leftrightarrow|g\rangle$
is off-resonant driven by a time-dependent microwave pulse with Rabi
frequency $\Omega_{L,j}(t)$ and the same detuning $\Delta_{j}$,
respectively. The interaction Hamiltonian with $\hbar=1$ for the
hybrid system is given by
\begin{eqnarray}    
H_{I}(t)=\sum_{j=1}^{2}\eta_{j}(t)\,a\sigma_{j}^{\dag}\,+H.c.,
\label{interaction}
\end{eqnarray}
where $\sigma_{j}^{\dag}=|e\rangle_j\langle g|$ and
$\eta_{j}(t)=g_{\!j}\,\Omega_{L,j}(t)/\Delta_{j}$ is the effective
coupling strength. Obviously, it is easy to realize full control
of $\eta_{j}(t)$ by changing Rabi frequency $\Omega_{L,j}(t)$ of the
microwave pulse when the parameters $g_{j}$ and $\Delta_{j}$ are
prescribed. The Hamiltonian $H_{I}(t)$ conserves the total excitation
number $N=\sum_{j=1}^{2}\sigma_{j}^{\dag}\sigma_{j}^{-}+n_c$ during
the dynamical evolution with $n_c$ being the photon number in the
resonator and $\sigma_{j}^{\dag}=(\sigma_{j}^{-})^{\dag}$. The whole
system evolves in the one-excited subspace spanned by
$\{|\phi_{1}\rangle=|0ge\rangle, |\phi_{2}\rangle=|0eg\rangle,
|\phi_{3}\rangle=|1gg\rangle\}_{c,1,2}$, where the subscripts ${c}$,
${1}$, and ${2}$ donate the resonator mode, the first NVE, and the
second NVE, respectively. In the basis of $\{|\phi_{1}\rangle,
|\phi_{2}\rangle, |\phi_{3}\rangle\}$, the interaction Hamiltonian
$H_{I}(t)$ is equivalent to the Hamiltonian $H_{0}(t)$. Consequently,
one can achieve the robust and accurate population transfer and
maximally entangled state generation between two NVEs, where the
cavity state is employed as an ancillary.

\begin{figure}[!h]
\begin{center}
\includegraphics[width=8cm,angle=0]{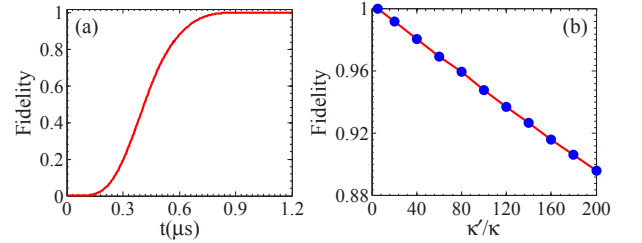}
\caption{(a) Fidelity of the population transfer
scheme with MSDs from $|\phi_{2}\rangle=|0eg\rangle$ to
$|\phi_{1}\rangle=|0ge\rangle$ under the influence of dissipations.
In the simulation, $\kappa^{-1}=50~\rm\mu s$ \cite{decay},
$\gamma^{-1}=6~\rm ms$, $\gamma_\varphi^{-1}=600~\rm\mu s$
\cite{dephasing}, and other parameters are the same as those for
Fig. \ref{fig1}(c).  (b) Fidelity of the population transfer scheme
with MSDs for different cavity decay rates $\kappa'$ (in units of
$\kappa$). }\label{fig4}
\end{center}
\end{figure}

In the presence of dissipations, the dynamics of the NVE-TLR hybrid
system is described by the Lindblad master equation:
\begin{eqnarray}    
\frac{d\rho}{dt}\!=\!-i\,[H(t),\rho]\!+\!\kappa D[a]\rho\!+\!\gamma
D[\sigma^{-}]\rho\!+\!\gamma_\varphi D[\sigma^{z}]\rho,
\end{eqnarray}
where $\rho$ is the density matrix operator for the hybrid system,
$H(t)$ is the Hamiltonian in the form of Eq.~(\ref{interaction}),
$D[L]\rho=(2L\rho L^{+}-L^{+}L\rho-\rho L^{+}L)/2$, $\kappa$ is the
decay rate of TLR, and $\gamma$ and $\gamma_\varphi$ are the
relaxation and dephasing rates of NVE, respectively. For the
proposed transfer scheme, the fidelity is defined as
$F=\langle\phi_{1}|\rho|\phi_{1}\rangle$ with $|\phi_{1}\rangle$
being the corresponding ideally final state under the population
transfer on its initial state $|\phi_{2}\rangle$. By choosing the
feasible experimental parameters as: $g/{2\pi}=20~\rm MHz$,
$\Delta/{2\pi}=200~\rm MHz$, $\kappa^{-1}=50~\rm\mu s$,
$\gamma^{-1}=6~\rm ms$, $\gamma_\varphi^{-1}=600~\rm\mu s$, and
$\Omega_{L,j}(t)=\Omega_{0}e^{-[(t-t_{j})/T]^2}~\rm MHz$ with
$t_{1}=0.75~\rm\mu s$, $t_{2}=0.25~\rm\mu s$, $T=0.408~\rm\mu s$,
$\Omega_{0}/{2\pi}=16~\rm MHz$, which meets the adiabatic condition
that $\int_{0}^{\tau}\Omega_{L,j}(t)dt\gg1$ with $\tau=1.2~\rm\mu s$
\cite{STIRAP}, one can find that the proposed scheme with MSDs can
realize perfect population transfer with the fidelity being $100\%$,
as shown in Fig. \ref{fig4}(a). To illustrate the robustness of the
present scheme, we also simulate the dependence of the fidelity $F$
versus the photon decay rate $\kappa'$ in Fig. \ref{fig4}(b). It shows
that a high fidelity of $89.60\%$ can still be obtained
even for $\kappa'/\kappa=200$. The reasons are two-manifold: The
cavity state is just used as an ancillary in the present scheme, so
it is insensitive to the photon decay in the resonator; the mean
photon number $\bar{n}=\langle a^{\dagger}a\rangle$, in consistency
with the population $P_{3}$ for intermediate state
$|\phi_{3}\rangle$ in Fig. \ref{fig1}, remains a trivial value
during the transfer process, which cannot achieve the complete
occupation of photon states \cite{photon}. The above results
suggest that time evolution of the populations with MSDs is more
robust against control parameter fluctuations and imperfections
than STIRAP.

\section{Discussion and Summary} \label{sec5}

We consider the feasibility with the current accessible parameters
in the NVE-TLR hybrid system. For an NVE placed at the antinodes of
the magnetic field of the full-wave mode of the TLR, the coupling
strength $g/{2\pi}=16~\rm MHz$ between them is reported
experimentally \cite{NVEs-TLR1,photon}. The amplitude of microwave
pulse is available with the current experiment parameter
$\Omega_{0}/{2\pi}=16~\rm MHz$ \cite{driving}. The detuning is
$\Delta/{2\pi}=160~\rm MHz$ so that $\Delta\gg g$ and $\Delta\gg
\Omega_{0}$, which can adiabatically eliminate the state
$|a\rangle$. From Eq.~(\ref{interaction}), the effective coupling
strength is $\eta_{j}/{2\pi}=1.6~\rm MHz$. When the coupling
strength $g$ and detuning $\Delta$ remain invariant, we have full
control of the population transfer and entanglement generation by
controlling flexibly the time-dependent Rabi frequency
$\Omega_{L,j}(t)$ of the microwave pulse with a single-shot
operation. The microwave coplanar waveguide resonators with the
decay rate of $\kappa^{-1}=50~\rm\mu s$ can be reached \cite{decay}.
The dephasing time of $T_{2}>600~\mu s$ for an NVE in bulk
high-purity diamond has been experimentally observed at room
temperature \cite{dephasing}. An optimized dynamical decoupling
microwave pulse has been demonstrated to increase the dephasing time
of NVE from $0.7~ms$ to $30~ms$ \cite{DD1}. Moreover, our
transitionless scheme with MSDs requires fewer resources, one TLR and two NVEs,
which greatly simplifies the experimental complexity.

In summary, we have presented a simple scheme for physically
feasible TQD for three-level quantum systems with MSDs, which is used
to realize perfect population transfer and entanglement generation
in a single-shot operation. Our
experimentally realizable transitionless protocol based on the
NVE-TLR hybrid system requires fewer physical resources and simple
procedures (one-step indeed), works in the dispersive regime, and is
robust against decoherence and control parameter fluctuations. These features
make our protocol more accurate for the  manipulation of the evolution of
three-level quantum systems than previous proposals, which may
open up further experimental realizations for robust quantum
information processing with MSDs.



\section*{ACKNOWLEDGMENTS}

We would like to thank Dr. Sof\'{\i}a Mart\'{\i}nez-Garaot and Wei Xiong for helpful
discussion. This work is supported by the National Natural Science
Foundation of China under Grants No. 11474026 and No. 11505007, and
the Fundamental Research Funds for the Central Universities under
Grant No. 2015KJJCA01.

\end{document}